\documentclass[twocolumn,prl,superscriptaddress,floatfix,showpacs]{revtex4-1}
\usepackage[utf8]{inputenc}
\usepackage{amsmath}
\usepackage{amssymb}
\usepackage{graphicx}

\makeatletter
\usepackage{graphics}
\usepackage{epsfig}
\usepackage{color}

\usepackage[normalem]{ulem}
\newcommand{\be}{\begin{equation}} \newcommand{\ee}{\end{equation}}
\newcommand{\bea}{\begin{eqnarray}} \newcommand{\eea}{\end{eqnarray}}

\makeatother

\begin{document}

\title{Kardar-Parisi-Zhang type dynamics with periodic tilt dependence of the propagation velocity in 1+1 dimensions}

\author{Peter Grassberger} \affiliation{JSC, FZ J\"ulich, D-52425 J\"ulich, Germany}

\date{\today}

\begin{abstract}
We consider the evolution of interfaces with a diffusive term and a generalized Kardar-Parisi-Zhang (KPZ) 
non-linearity, which results in a propagation velocity that depends periodically on the tilt of the interface. 
Using large scale simulations of a model class with these properties in 1+1 dimensions, we show that the 
fluctuations are in general still in the KPZ universality class, but a new universality class seems to appear 
in the limit of weak non-linearity. We argue that this is the typical behavior of any interface model with 
periodic tilt dependence.
\end{abstract}
\maketitle

Interfaces are everywhere. And in general non-equilibrium situations, they usually move and fluctuate 
\cite{Family,Meakin,Barabasi}. In the present letter, we will only consider one-dimensional interfaces in 
two-dimensional media. Maybe the most simple model for a moving and fluctuating interface is the one described 
by the Edwards-Wilkinson (EW) equation  \cite{edwards}
\be
   \frac{\partial h(x,t)}{\partial t} = v + \nu \nabla^2 h(x,t) + \eta(x,t).
\ee
Here, $h(x,t)$ is the height of the interface (assumed to be a univalued continuous function, $v$ is the 
propagation velocity of flat interfaces in the absence of noise, and $\eta(x,t)$ is Gaussian white noise. 
Due to the latter, the Laplacian of $h$ is strictly spoken not well defined, but this can be easily cured. 
Although the same is much less obvious for the next more complicated model, the Kardar-Parisi-Zhang
(KPZ) equation \cite{KPZ}
\be
   \frac{\partial h(x,t)}{\partial t} = v + \nu \nabla^2 h(x,t) + \frac{\lambda}{2}(\nabla h)^2 + \eta(x,t),
\ee
also in this case all derivatives can be interpreted in a mathematically rigorous way \cite{hairer}. 

One important difference between the EW and KPZ models is that the former applies to interfaces in equilibria,
while the latter is a genuinely non-equilibrium model, without detailed balance. Also, solving the EW equation
is simple by means of Fourier transformation (the EW equation is linear), while this is not at all true for KPZ. 
Indeed, the KPZ equation in higher dimensions is still not well understood, but rigorous studies in 1+1 dimensions
have brought us closer to a complete understanding in recent years \cite{Spohn,Takeuchi}.

Most of the complications of the KPZ model as compared to the EW model follow from the fact that the propagation 
velocity of a tilted straight interface depends on the tilt angle, which is not true for EW. Assume an isotropic
medium, and a straight horizontal interface that moves (in the absence of noise) upward with velocity $v$. In 
a rotated frame, where the axes make an angle $\phi$ with the original ones, the interface is tilted, but can be 
still viewed as moving up, in the direction of the new $z$-axis. But the velocity will then be $v/\cos(\phi)$ 
instead of $v$. 

For curved interfaces this has additional consequences: A hat-convex interface moving upward 
will widen, while a cup-convex interface with become narrower, until it finally develops a cusp singularity.
For the full KPZ model with noise -- and with non-linearity $\lambda$ as dictated by rotational invariance in
isotropic media -- this implies that typical interfaces are up-down asymmetric. Interfaces moving up consist
essentially of (noisy) cap-convex arcs, while downwards moving interfaces are chains of cup-convex arcs.

More general behavior can result from non-trivial anisotropy of the medium and of the microscopic mechanism.
Phenomenologically this can be modelled by replacing the quadratic term $\lambda(\nabla h)^2/2$ by some other 
non-linear function $G(\nabla h)$, and a corresponding velocity dependence. One expects KPZ scaling 
whenever $G(x)$ has a quadratic minimum (or maximum) at $x=0$, while EW scaling is expected when $G(x)$ is flat 
in an entire interval $-\epsilon < x < \epsilon$. More interesting is the case where $G''(x)=0$ just at $x=0$. 
The specific case where $G(x)$ has an inflection point at $x=0$ was studied in \cite{Devillard}, where EW 
scaling was found with possible logarithmic corrections.

In the present letter, we shall study another special case of non-linearity where the mean velocity of growth
is periodic in the `tilt' $a$,
\be
   v(a) = v(a+1), \qquad {\rm where} \;\; a = \tan(\phi).
\ee

To be precise, we study the following model: We have a 1-d lattice of even size $L$ and with discrete time 
$t=0,1,\ldots$. An integer variable $n(x,t)$ is defined on each even site at even times, and on odd sites at 
odd times. Initial conditions are tilted periodic, $n(x+L,0)=n(x,0)+aL$, with integer $aL$. The dynamics is 
synchronous, with the updating rule
\be
    n(x,t+1) = \frac{1}{2} n_{\rm in}(x,t+1) + \eta (n_{\rm in}(x,t+1))  \label{eq:1d}
\ee
where
\be
    n_{\rm in}(x,t+1)= n(x-1,t) +   n(x+1,t).
\ee
Here, $\eta (n_{\rm in})$ is the simplest non-trivial noise term that depends only locally on $n_{\rm in}$ and 
preserves the integer nature of $n(x,t)$ \cite{DMG},
\bea
    \eta(n_{\rm in}) = \left\{ \begin{array}{r@{\quad}l@{\quad}l}
                      \pm\frac12 & \mbox{with prob's} \;(q,1-q) \;\mbox{for odd}\;  n_{\rm in}  \\
                      \pm 1,0   & \mbox{with prob's} \;(p_{\pm 1}, p_0)\;\mbox{for even}\; n_{\rm in}  \\
                 \end{array} \right.    \label{noise}
\eea
where $p_{-1}+ p_0+ p_1 = 1$.

{\bf Theorem:} {\it  For this model, the velocity depends periodically on the tilt with period 1, i.e.
$v(a) = v(a+1)$.}

{\bf Proof:} The instantaneous propagation velocity $v(a,t)$ of an interface ${\cal I}_t = \{n(x,t)\}$ at 
time $t$ is defined as 
\be
    v(a,t) = \frac{2}{L} \sum_x [\langle n(x,t+1) \rangle - n(x,t)],
\ee
and $v(a)$ is just its forward time average.  Due to Eq.(\ref{eq:1d}), we have 
\bea
   v(a,t) &=& \frac{2}{L} \sum_x \langle \eta (n_{\rm in}(x,t+1)) \rangle \label{m} \nonumber \\
	  &=& \frac{2}{L} [(q-1/2)\langle m_{\rm odd}(t+1)\rangle   \\
	  & & + (p_1-p_{-1}) \langle m_{\rm even}(t+1)\rangle], \nonumber 
\eea
where $m_{\rm odd}$ and $m_{\rm even}$ are the numbers of sites where $n_{\rm in}$ is even and odd, 
respectively.

    Assume now that $n'(x,t) = n(x,t)+x$. If $n(x,t)$ has average tilt $a$, then
$n'(x,t)$ has average tilt $a' = a+1$. On the other hand, both have the same instantaneous (and thus 
also average) velocities, because even-ness and oddness are not changed when going from $n_{\rm in}(x,t)$ 
to $n'_{\rm in}(x,t)$ and, since $\eta'=\eta$, also $n'(x,t+1) = n(x,t+1)+x.\;\;$ QED.

\begin{figure}
	\vglue -.2cm
\begin{centering}
\includegraphics[scale=0.28]{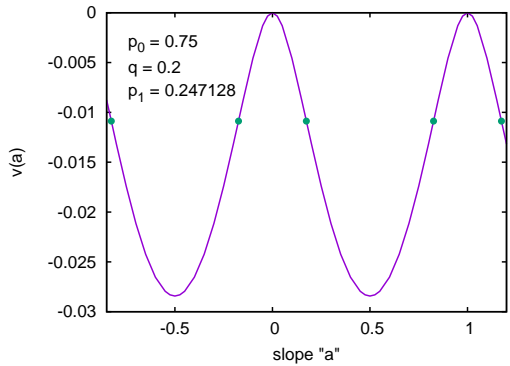}
\vglue -.6cm
\par\end{centering}
\caption{\label{fig1} (color online) Average asymptotic interface velocity for tilted interfaces
   with tilt $a$. Parameters are such that interfaces without tilt are critical, i.e. $v(0)=0$.
   The curve resembles a cosine, but is clearly distinct from it. The heavy dots indicate
   inflection points where the curvature is zero.}
\end{figure}

To illustrate this, $v(a)$ is plotted in Fig.~1 against $a$ for one particular set of control 
parameters. These are such that flat interfaces are `critical', $v(0)=0$, and slightly tilted ones move down, 
$v''(0) < 0$. Therefore, in this case, typical globally flat interfaces consist of chains of cup-convex arcs, 
and we expect that interfaces that are strictly flat at time $t=0$ (having $n(x,0)=0$ for all even $x$) move 
{\it up} during a transient,
\be
   v(0,t) \sim c\; t^{-2/3}       \label{v-KPZ}
\ee
with $c > 0$ \cite{KPZ}, such that the average height of the interface should scale as
\be
   \langle n(t)\rangle =  \frac{3c}{2} t^{1/3} + const.   \label{n-KPZ}
\ee
Finally, its width should scale as \cite{KPZ}
\be
    W^2(L,t) \equiv {\rm Var}[n(x,t)] \sim L^{2\alpha} f(t/L^z)   \label{W-KPZ}
\ee
with $f(u)\sim 1$ for $u\gg 1, f(u)\sim u^{2\beta}$ for $u\ll 1, \alpha = 1/2, \beta = 1/3$, and $z = 3/2$.

In the following we shall, for simplicity, stick to the critical case, i.e. we shall assume that $v(0)=0$ (the 
situation is basically the same for $v(0)\neq 0$, but the discussion is more complicated). One then 
finds that $v''(0) < 0$ whenever $q<1/2$ (as in Fig.~1), while $v''(0) > 0$ if $q>1/2$. Indeed, for $q=1/2$
the model is critical for $p_1=p_{-1}$, and is in the EW universality class due to up-down symmetry. In that 
case, $\langle n(t)\rangle = const,\; \alpha = 1/2,\; \beta = 1/4$, and $z=2$.

\begin{figure}
\begin{centering}
\includegraphics[scale=0.27]{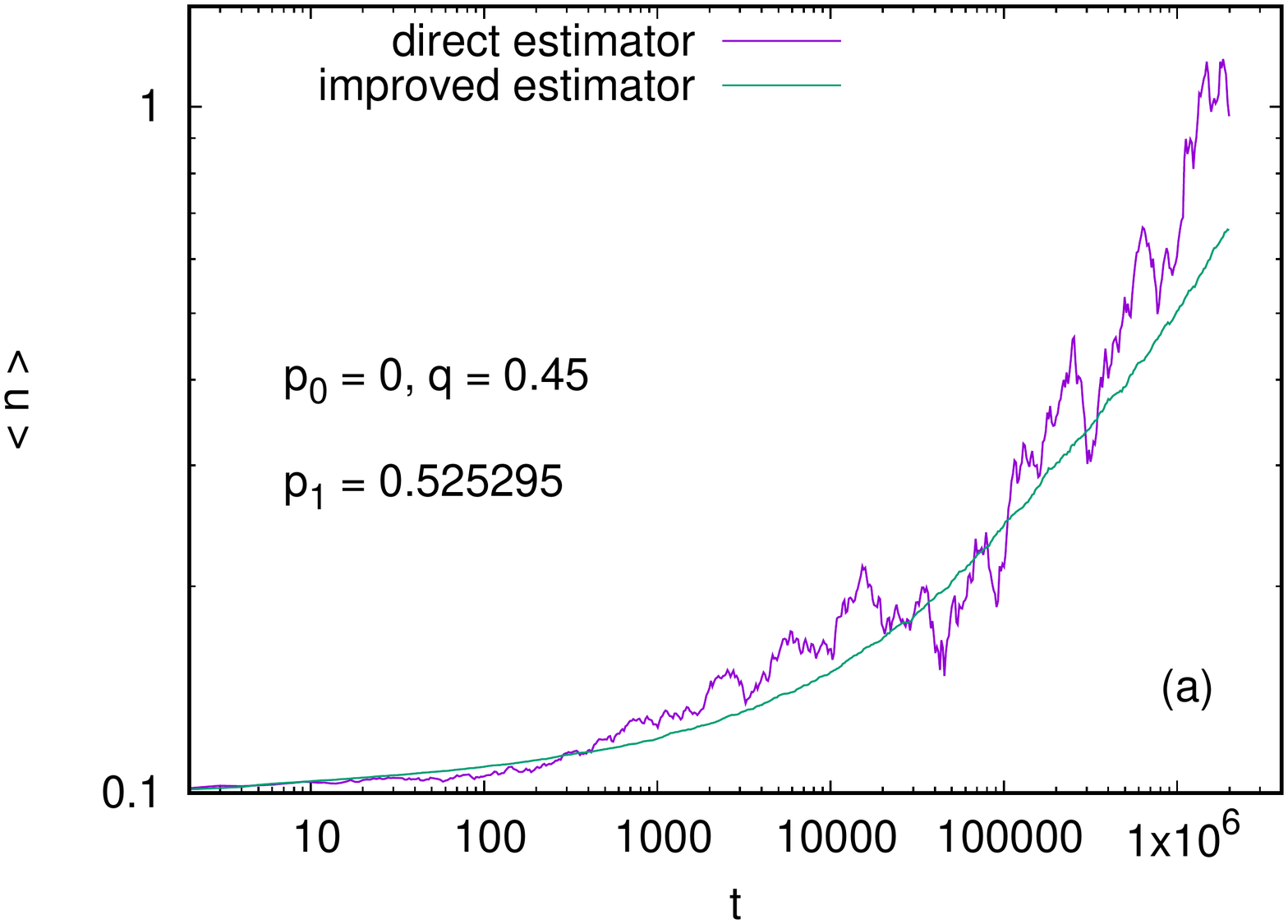}
\vglue -.7cm
\includegraphics[scale=0.27]{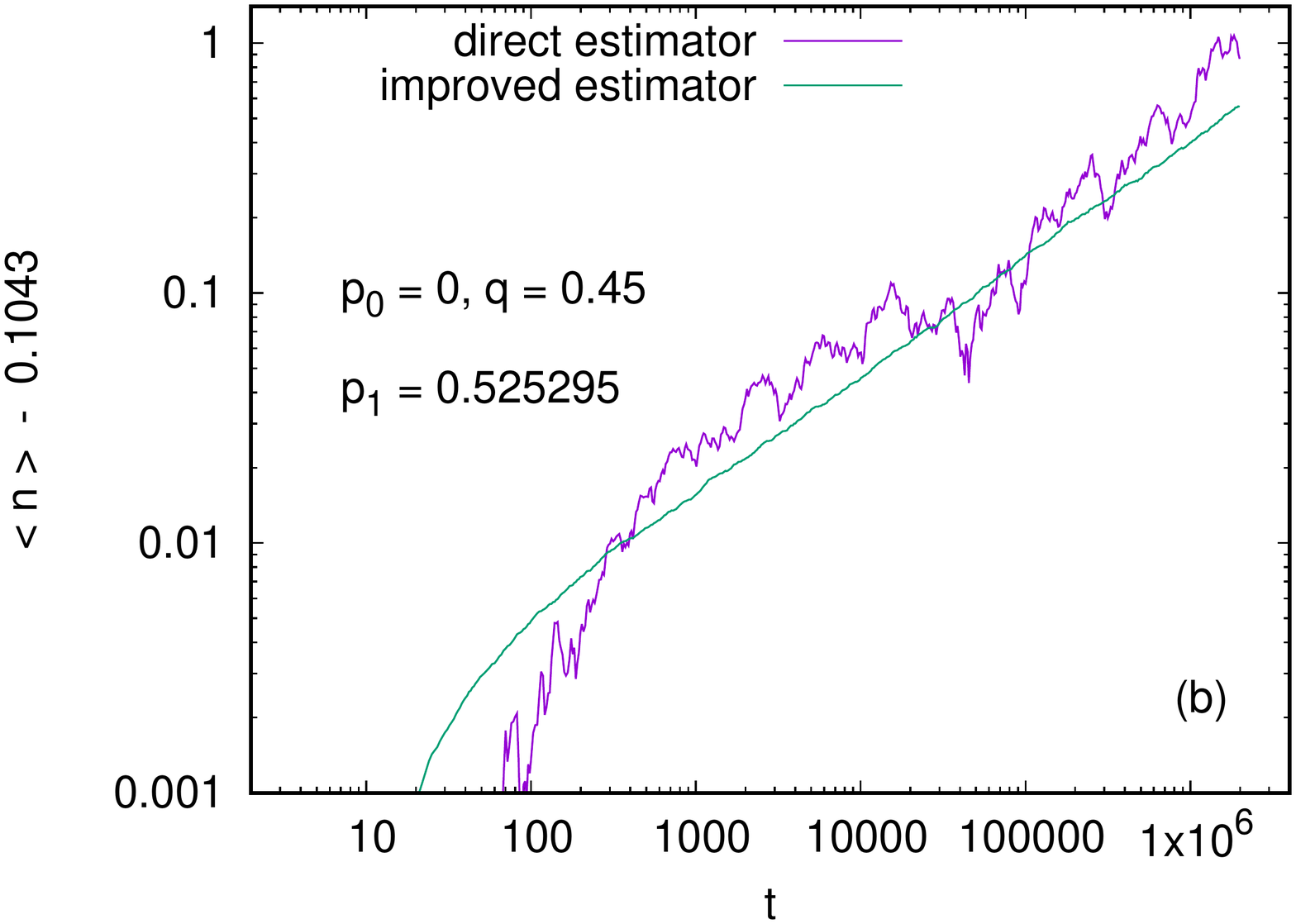}
\vglue -.5cm
\par\end{centering}
\caption{\label{fig2} (color online) Average interface height for initially flat interfaces
   starting at height $n(t=0) = 0$.
   Parameters $[ p_0 = 0, q = 0.45, p_1 = 0.525295]$ 
   are again such that interfaces are critical, i.e. $v(0)=0$. Panel (a) shows a log-log plot of 
   $\langle n(t)\rangle$ versus $t$, while panel (b) shows $\langle n(t)\rangle - c$, where the 
   constant $c$ is chosen such that the scaling region with $\langle n(t)\rangle - c \sim t^x$ is maximal, 
   with the exponent $x$ fitted together with $c$. In both panels, the improved estimator (smooth curve) 
   is compared to the much more noisy direct measurement.}
\end{figure}

In order to verify Eqs.~(\ref{v-KPZ}-\ref{W-KPZ}) numerically, we use large scale simulations. In these 
simulations we can either measure $n(t)$ directly, or we can measure $m_{\rm odd}$ and $m_{\rm even}$ and use 
Eq.~(\ref{m}). The latter is always more precise (using it, we are not affected by random number fluctuations
when actual up/down steps are made guided by $m_{\rm odd}$  and $m_{\rm even}$ \cite{footnote1} 
), but the difference depends grossly on the control parameters.
In some cases variances are reduced roughly by a factor 2 when using Eq.~(\ref{m}), but when $q$ is close to 
$1/2$ the reduction is dramatic (see Fig.~2; variances are reduced by more than 2 orders of magnitude). Indeed, 
the present analysis would have been impossible using the direct measurement of $\langle n(t)\rangle$.

In the following we shall only show results for $ t \ll L^{3/2}$ (with $L=2^{19}$), 
in which case the variance is independent of $L$ and Eq.~(\ref{W-KPZ}) simplifies to 
\bea
   {\rm Var}[n(x,t)]\sim t^{2\beta} \sim \left\{ \begin{array}{r@{\quad}l@{\quad}l}
	                                     t^{2/3} & \;\mbox{for } q\neq 1/2\\
                                             t^{1/2} & \;\mbox{for } q = 1/2\;.\\
                                       \end{array} \right.    \label{W_largeL}
\eea
Finally, we shall always choose the parameters such that $v(0)=0$. This then leaves us with two control parameters,
for which we can use e.g. $p_0$ and $p_1-p_{-1}$, or $p_0$ and $q$. For fixed $p_0$, the difference $p_1-p_{-1}$
decreases monotonically with $q$, being minimal for $q=1$ and maximal for $q=0$ (see Fig.~3). Due to the up-down
symmetry of Fig.~3, we shall in the following restrict our discussion to $0\leq q \leq 1/2$.

\begin{figure}
	\vglue -.6cm
\begin{center}
\includegraphics[scale=0.30]{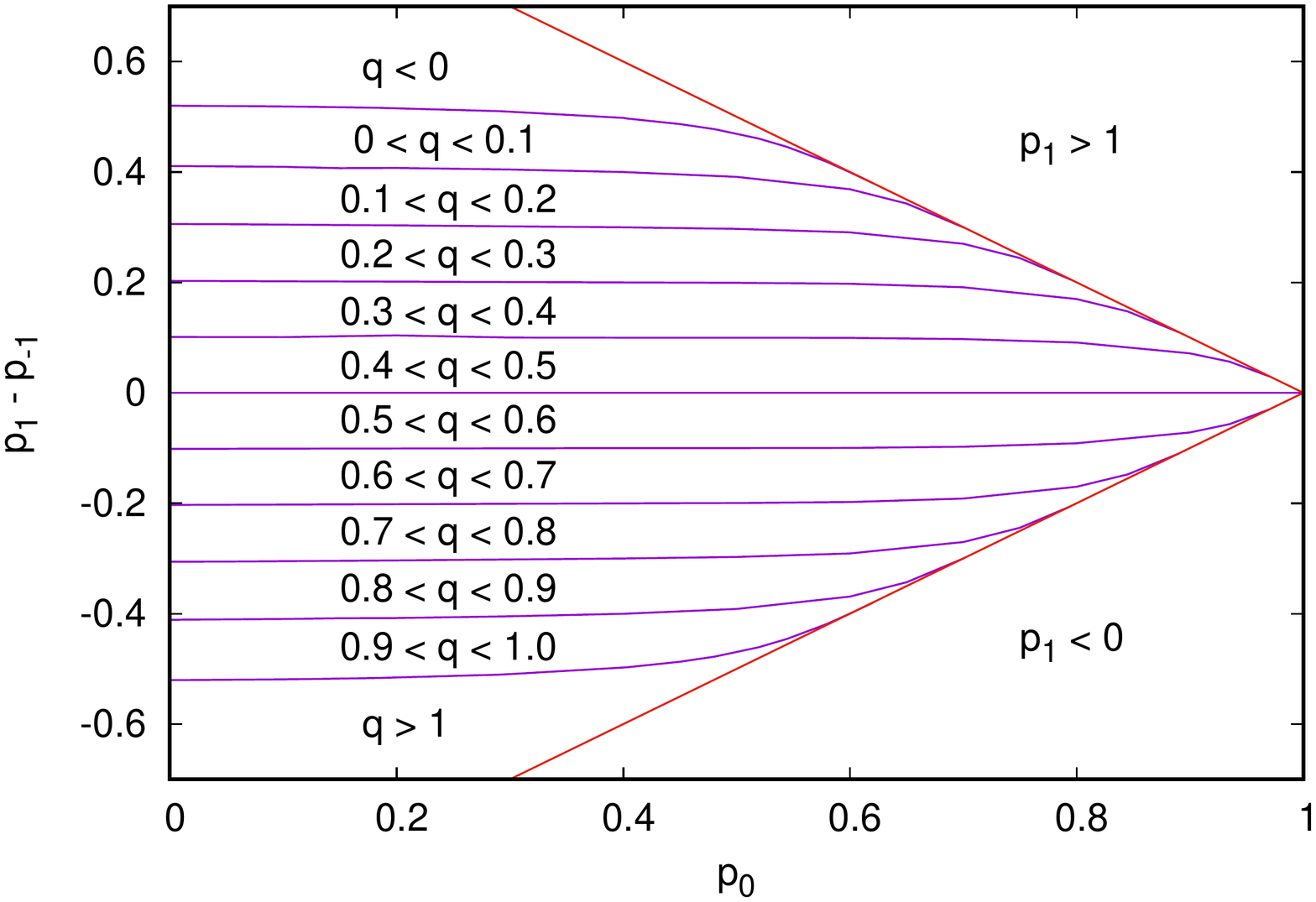}
\end{center}
\vglue -.8cm
\caption{(color online) Contour lines $q = 0,0.1,0.2,...1.0$, plotted with $p_0$ on the $x-$axis versus $p_1-p_{-1}$ 
	on the $y-$axis. The triangular regions at the right (with $p_1<0$ or $p_1>1$) are unphysical, as are the 
	regions with $q>1$ and $q<0$. EW scaling is expected at $q=1/2$, and KPZ everywhere else.}
\label{fig:contour}
\end{figure}

\begin{figure}
\begin{center}
\includegraphics[scale=0.30]{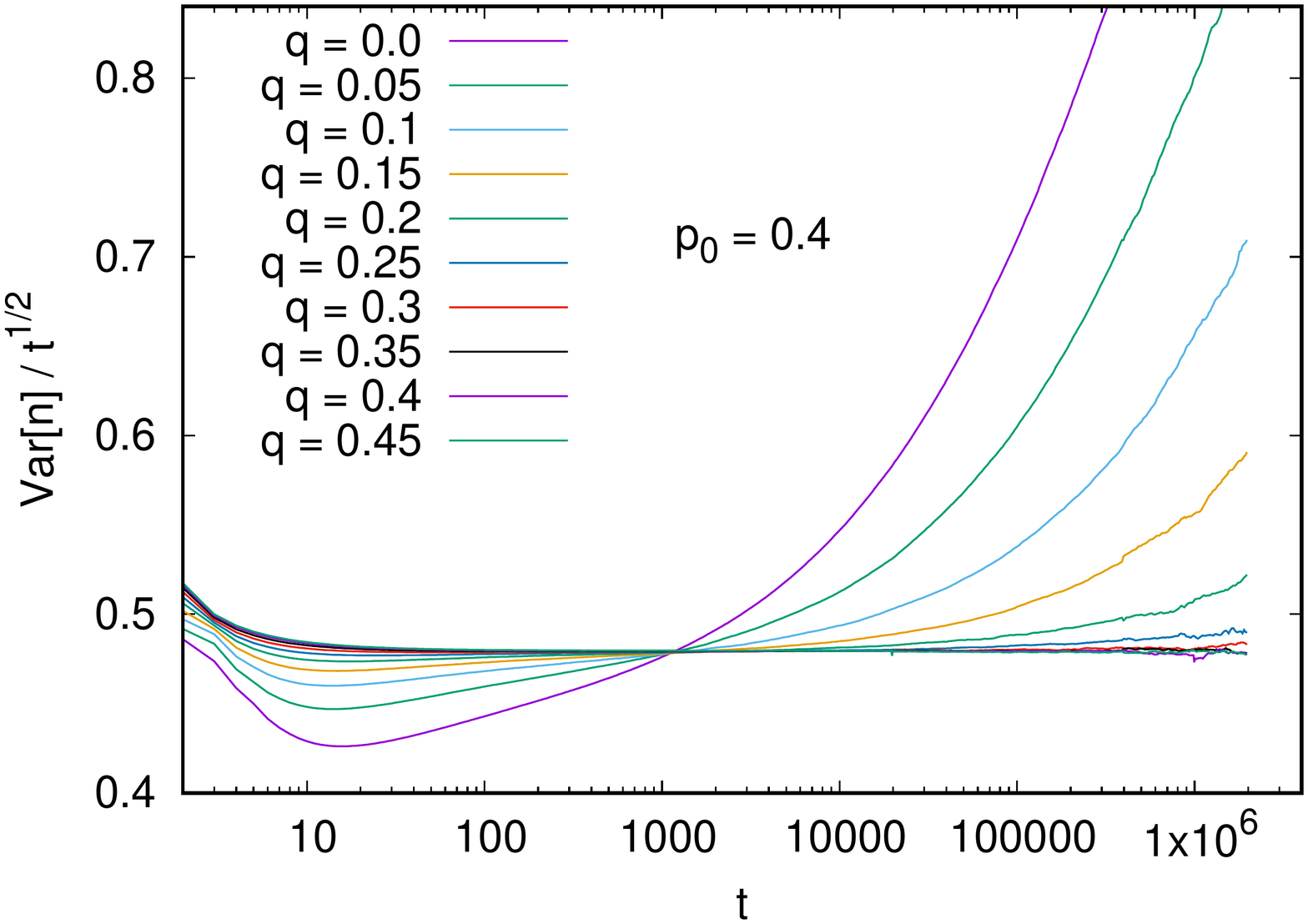}
\end{center}
\vglue -.8cm
\caption{(color online) Log-log plots of $t^{-1/2}\;{\rm Var}[n(x,t)]$ against $t$ for $p_0=0.4$ and for ten values
        of $q$.}
\label{fig:skew-b_log}
\end{figure}

\begin{figure}
\begin{center}
\includegraphics[scale=0.30]{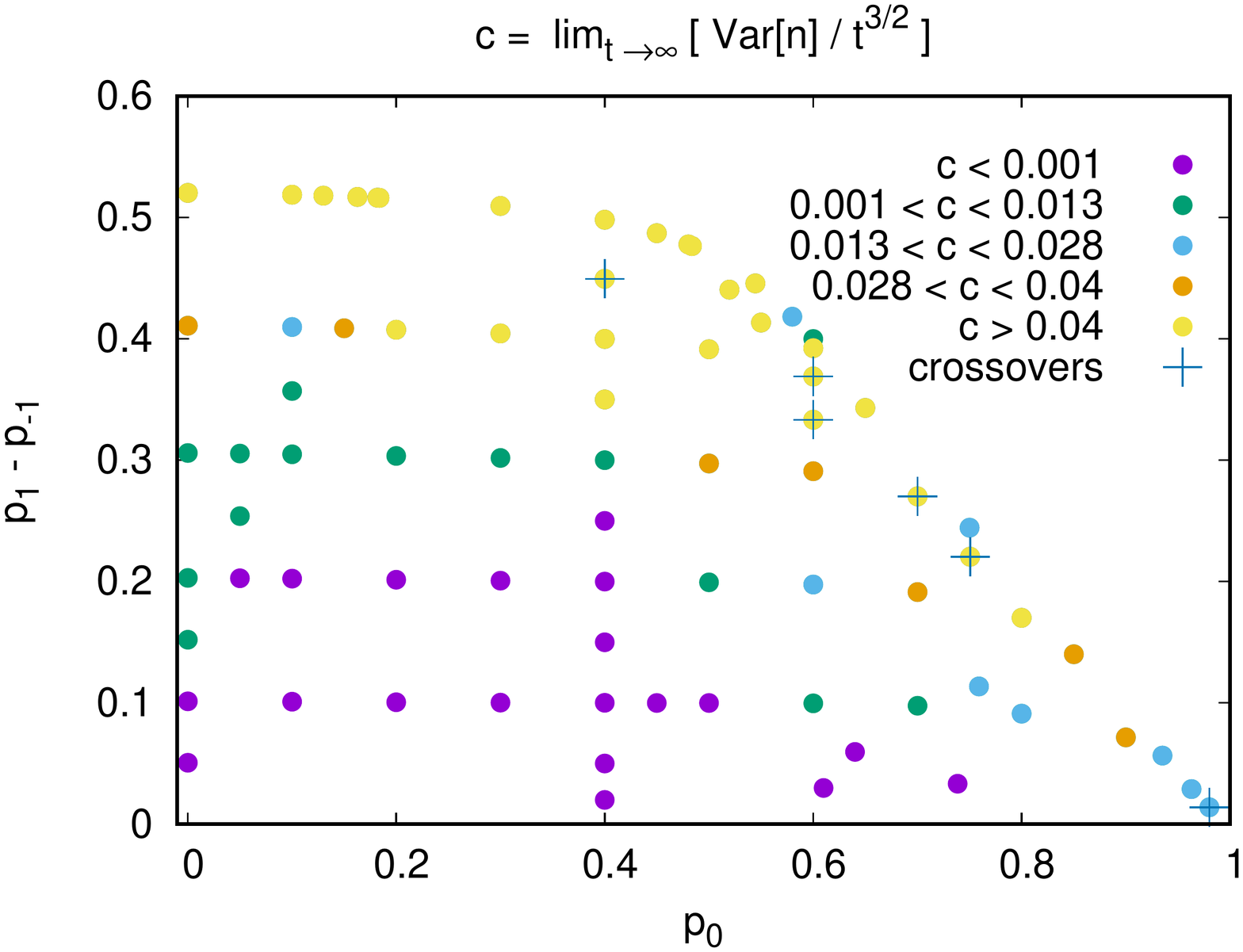}
\end{center}
\vglue -.9cm
\caption{(color online) Amplitudes $c = \lim_{t\to\infty}{\rm Var}[n(x,t)]/t^{2/3} $ of the expected KPZ scaling law,
	plotted against the two control parameters $p_0$ and $p_1-p_{-1}$. In a wide region (magenta and green 
	points) $c$ is compatible with zero. Crosses indicate clear cross-overs.} 
\label{fig:variances-b}
\end{figure}

We shall first discuss variances of $n(x,t)$, since we do not need for them precise estimates of `critical' 
parameter values, while such precise estimates are needed in discussing $\langle n(t)\rangle$.
In Fig.~\ref{fig:skew-b_log} we show ${\rm Var}[n(x,t)]/t^{1/2}$ versus $t$, for $p_0=0.4$ and for ten values 
of $q$. All curves for $q \leq 0.25$ bend upwards for large $t$, indicating thereby that we do not see EW scaling.
But for no value of $q$ the rise of ${\rm Var}[n(x,t)]/t^{1/2}$ is as strong as we would expect from pure KPZ
scaling, indicating that we are in a cross-over regime. For $q \geq 0.3$, however, the curves seem to be perfectly 
flat for large $t$, suggesting very clean EW scaling. But we cannot rule out, of course, that at least some 
of these curve would bend uppwards for even larger $t$ -- and observing the trend for $q \leq 0.25$, this looks
indeed very plausible. Thus we conclude that KPZ is observed for small $q$, while there is a cross-over to another 
universality class -- maybe EW -- either at $q=1/2$ or at a critical value $q_c < 1/2$. Similar results for 
other values of $p_0$ are summarized in Fig.~\ref{fig:variances-b}, where we see that 
${\rm Var}[n(x,t)]/t^{2/3}$ seems to tend towards zero in a very large part of control parameter space.

\begin{figure}
\begin{center}
\includegraphics[scale=0.30]{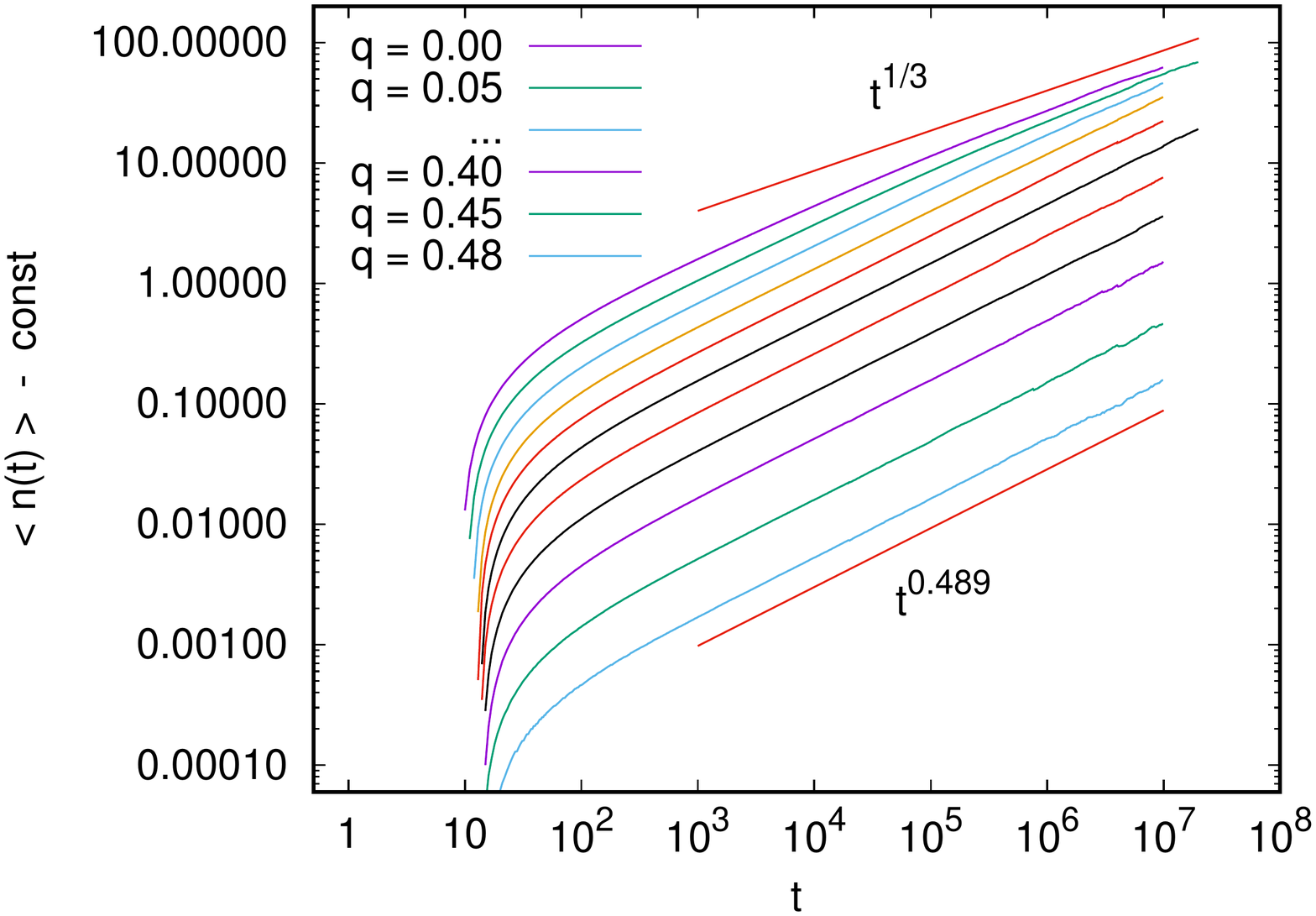}
\end{center}
\vglue -0.9cm
\caption{(color online) Log-log plots of $\langle n(t)\rangle - const$ against $t$ for $p_0=0.4$. The values of 
	$q$ the same ten
	as in Fig.~\ref{fig:skew-b_log}, in addition to $q=0.48$. The uppermost curve is for $q=0$. For each $q$, 
	the constant is fitted such that the curve becomes, for large $t$, closest to a straight line. Also shown
	are the straight lines corresponding to pure power laws $t^x$ with $x=\beta=1/3$ (KPZ; upper line) and 
	$x \approx 1/2$ (``pseudo-EW"; lower line). The exponent $0.489$ of the latter is the best overall fit,
	but we propose that the exact value is $0.50\pm 0.2$.}
\label{fig:m_aver-4}
\end{figure}

Let us now discuss the scaling of $\langle n(t)\rangle$. Precise parameter values where $v(0)=0$ are obtained 
as in Fig.~2b by making log-log plots of $\langle n(t)\rangle - const$ versus $t$, where the constant is chosen 
such that the plots show the cleanest possible straight lines for large $t$. A priori we would have expected 
KPZ scaling for all $q\neq 1/2$, i.e. $\langle n(t)\rangle - const \sim t^{1/3}$, while the above scaling of 
variances would suggest $\langle n(t)\rangle = const$ for $q\approx 1/2$. Neither is seen in 
Fig.~\ref{fig:m_aver-4}, where such plots are shown, again for $p_0 = 0.4$. Nothing resembling EW scaling is 
seen for any value of $q$. Rather, for all $q>0.2$ the curves are much closer to $\langle n(t)\rangle - 
const \sim b t^{1/2}$ than to KPZ scaling. We shall call this ``pseudo-EW" scaling. As far as we know there is 
no known generalization of proper EW where $\langle n(t)\rangle \sim b t^{1/2} + O(1)$. Clear deviations from 
pseudo-EW towards KPZ are only seen for $q<0.2$. Indeed, even for $q=0$ we see no clean KPZ scaling, but we 
see a clear cross-over which suggests that KPZ scaling would hold asymptotically. Similar results are seen 
also for all other values of $p_0$. For $q=1/2-\epsilon$ and $\epsilon \to 0$, the amplitude $b$ scales roughly 
as $b\sim \epsilon$.

It could well be that -- as suggested by the last plots -- there is indeed a phase transition between KPZ 
scaling and a new ``pseudo-EW" universality class at values of $q < 1/2$. But we believe that 
this is not really the case. Rather we suggest that this transition occurs at $q=1/2$, i.e. for $q\approx 1/2$
there are extremely slow cross-overs which would be impossible to see with present-day computers. In any case,
there is a new universality class.

In addition, we also looked at the widths and average speeds of tilted interfaces. As expected, we found EW 
scaling at and only at inflection points of $v(a)$ \cite{Devillard}. Thus it is indeed only the curvature $v''(a)$ 
and not the linear dependence $v'(a)$ which determines whether we see EW scaling or not. But, in contrast to 
conventional wisdom and in agreement with the above findings, we do not see KPZ scaling when we approach the 
inflection point, i.e. when $v''(a)\to 0$. Rather we see again pseudo-EW scaling with leading behaviors
$\langle n(t)\rangle - const \sim t^{1/2}$ and ${\rm Var}[n(x,t)] \sim t^{1/2}$.

The intuitive reason for this transition to a new universality class -- either for some very small but non-zero
value of $v''(a)$, or in the limit $v''(a)\to 0$  -- 
is the following. As we said in the introduction, a 1-dimensional KPZ interface
consists, for large times $t$, essentially of chains of noisy arcs. As time goes on, these arcs keep their 
curvatures, but they become wider and thus also deeper. But for the present model this scenario cannot go
on forever. As the arcs become deeper, they become also steeper at their wings. Thus sooner or later they would
typically have slopes $>1$. But before this could happen, a point would be encountered where
interfaces tilted with this slope feel the opposite curvature of $v(a)$, and thus the arcs would stop
becoming steeper. It remains an open challenge to verify this scenario mathematically. It is also not yet clear 
how it generalizes to higher dimensions. Finally, the above heuristic arguments would suggest that the same 
novel universality class would not only be seen for periodic $v(a)$ dependence, but also when $v(a)$ is a shallow
``Mexican hat" function. Whether this is indeed true is another open question.

Acknowledgment: I deeply want to thank Deepak Dhar and Pradeep Mohanty for numerous discussions. Indeed, this work 
was performed as part of a larger joint project with them, the main part of which is still to be published. I am 
also much indebted to Jan Meinke for help in vectorizing the algorithm.


\begin{thebibliography}{99}
\bibitem{footnote1} This is an instance of variance reduction due to avoidance of random decisions. For other 
	applications of this strategy see, e.g., \cite{grassberger_percol_high-d,grassberger_log4d,Foster,JfWang}.
\end{thebibliography}
\end{document}